\input harvmac
\def\ra{\rightarrow}
\def\gsim{{~\raise.15em\hbox{$>$}\kern-.85em
          \lower.35em\hbox{$\sim$}~}}
\def\lsim{{~\raise.15em\hbox{$<$}\kern-.85em
          \lower.35em\hbox{$\sim$}~}}

\noblackbox
\baselineskip 14pt plus 2pt minus 2pt
\Title{\vbox{\baselineskip12pt
\hbox{hep-ph/9911370}
\hbox{IASSNS--HEP--99--106}
\hbox{WIS--99/36/Nov--DPP}
}}
{\vbox{
\centerline{Neutrino Parameters, Abelian Flavor Symmetries,}
\vskip .2cm
\centerline{and Charged Lepton Flavor Violation}
  }}
\centerline{Jonathan L. Feng$^a$, Yosef Nir$^{a,b}$ and
Yael Shadmi$^b$\foot{On leave of absence from Princeton University.}}
\medskip
\centerline{\it $^a$School of Natural Science,
Institute for Advanced Study}
\centerline{\it Princeton, NJ 08540, USA}
\centerline{feng,nir@ias.edu}
\medskip
\centerline{\it $^b$Department of Particle Physics}
\centerline{\it Weizmann Institute of Science, Rehovot 76100, Israel}
\centerline{yshadmi@wicc.weizmann.ac.il}
\bigskip

\baselineskip 18pt
\noindent

Neutrino masses and mixings have important implications for models of
fermion masses, and, most directly, for the charged lepton sector.  We
consider supersymmetric Abelian flavor models, where neutrino mass
parameters are related to those of charged leptons and sleptons. We
show that processes such as $\tau\ra\mu\gamma$, $\mu\ra e\gamma$ and
$\mu-e$ conversion provide interesting probes.  In particular, some
existing models are excluded by current bounds, while many others
predict rates within reach of proposed near future experiments. We
also construct models in which the predicted rates for charged lepton
flavor violation are below even the proposed experimental
sensitivities, but argue that such models necessarily involve loss of
predictive power.

\Date{11/99}

\newsec{Introduction}

The most problematic aspect of the Standard Model is the unnaturally
small ratio between the electroweak breaking scale and the Planck
scale, that is, the fine-tuning problem. Supersymmetry protects this
ratio against radiative corrections. Another puzzling aspect of the
Standard Model is the unexplained hierarchy in the fermion masses and
mixings, that is, the flavor puzzle. Approximate horizontal symmetries
give rise to selection rules that can account for this hierarchy. The
framework of {\it supersymmetric flavor models}, which combines these
two extensions of the Standard Model, is particularly interesting
because there is an interplay between the two ingredients.
Supersymmetry affects the flavor parameters since it requires the
Yukawa parameters to be holomorphic. Horizontal symmetries affect the
supersymmetry breaking parameters since these parameters are subject
to appropriate selection rules. One of the most attractive features of
supersymmetric flavor models is that, as a result of this interplay,
the measured values of fermion masses and mixings have implications
for supersymmetric contributions to flavor changing neutral current
processes.  Conversely, measurements of rare processes provide
possibly stringent tests of models with horizontal symmetries.

Recent measurements of the fluxes of atmospheric
\ref\ANSK{Y.~Fukuda {\it et al.}, the Super-Kamiokande collaboration,
 Phys.~Rev.~Lett.~81 (1998) 1562, hep-ex/9807003.}\
and solar
\ref\BKS{For a recent analysis, see, {\it e.g.},
 J.~N.~Bahcall, P.~I.~Krastev and A.~Yu.~Smirnov, Phys. Rev. D58 (1998)
 096016, hep-ph/9807216.}\
neutrinos have added to our knowledge of neutrino parameters. The
simplest interpretation of the experimental results concerning
atmospheric neutrinos (AN) is in terms of $\nu_\mu-\nu_\tau$
oscillations, with the following central values for the mass-squared
difference and mixing angle:
\eqn\ANpar{\Delta m^2_{23}\sim2\times10^{-3}\ {\rm eV}^2,\ \ \
\sin^22\theta_{23}\sim1.}
The simplest interpretation of the experimental results concerning
solar neutrinos (SN) is in terms of $\nu_e-\nu_x$ ($x=\mu$ or $\tau$)
oscillations, with one of the following sets of parameters:
\eqn\SNpar{
\matrix{&& \Delta m^2_{1x}\ [{\rm eV}^2] & \sin^22\theta_{1x} \cr &&&\cr
{\rm MSW(SA)}      && 5\times10^{-6}  &  0.006             \cr
{\rm MSW(LA)}      && 2\times10^{-5}  &  0.8               \cr
{\rm VO}           && 8\times10^{-11} &  0.8               \cr}}
Here MSW (VO) refers to matter-enhanced (vacuum) oscillations, and SA
(LA) stands for a small (large) mixing angle. These neutrino
parameters, together with the masses of the charged leptons,
\eqn\chamas{m_e\simeq0.51\ {\rm MeV},\ \ \ m_\mu\simeq106\ {\rm MeV},\ \ \
m_\tau\simeq1777\ {\rm MeV},}
constrain model building with horizontal symmetries in the lepton
sector.

While any measurement of neutrino parameters provides welcome guidance
for attempts to explain the fermion masses, the parameters of
Eqs.~\ANpar\ and \SNpar\ are particularly provocative.  In the
simplest realizations of models with horizontal symmetries, states
with large mixing must have similar masses.  In contrast, \ANpar\ and
\SNpar\ suggest both a large 2-3 mixing and a hierarchically 
suppressed $m_2/m_3$ ratio in the neutrino sector. (Here we assume
that the mass-squared differences are of order of the larger
mass-squared involved.) The experimental data has thus motivated many
studies of extensions of, or alternatives to, the simplest models.
The possibilities explored include:
\item{(i)} The neutrino masses arise from different sources.  For
example, in the framework of supersymmetry without $R$-parity, the
heaviest neutrino acquires its mass at tree level while the lighter
ones become massive only through loop effects.
\item{(ii)} Certain Yukawa couplings vanish because of holomorphy
(`holomorphic zeros').
\item{(iii)} The horizontal symmetry is discrete.
\item{(iv)} The horizontal symmetry is broken by two small breaking
parameters of equal magnitude and opposite charge.
\item{(v)} The hierarchy in masses is accidental.

While there are clearly many possibilities, most of these models solve
the `large mixing--large hierarchy' problem in the following way: the
neutrino mass hierarchy follows from the structure of the neutrino
mass matrix, while the large mixing arises from diagonalizing the
charged lepton mass matrix.  With the standard model fermion content
alone, this approach produces the desired neutrino properties with no
other experimentally interesting implications. For example, although
the mixing of neutrinos induces, at the loop-level, flavor mixing in
the charged leptons, it is at an unobservably small level: for
neutrino masses of 100 eV, $B(\mu\ra e\gamma) < 10^{-40}$
\ref\chengli{See, {\it e.g.}, T.-P.~Cheng and L.-F.~Li, {\it Gauge
Theory of Elementary Particle Physics} (Oxford University Press, Great
Britain, 1984).}.

When extended to {\it supersymmetric} models, however, the large
mixing in the charged lepton mass matrix has profound experimental
implications. Supersymmetry introduces additional scalars that are
governed by the same horizontal symmetries.  As we shall see, these
scalars can induce charged lepton flavor violation (LFV) at current
experimental sensitivities. For example, the large mixing angle
suggested by AN implies a rate for $\tau \ra\mu\gamma$ that is
typically close to the present bound if slepton masses are around
100~GeV. Thus, the new results on neutrino parameters warrant a close
analysis of charged LFV in the context of supersymmetric flavor
models.

At present, searches for charged LFV have yielded only upper bounds.
Among the most stringent are the bounds on radiative decay
\nref\meexp{M.~L.~Brooks {\it et al.}, MEGA Collaboration,
 Phys. Rev. Lett. 83 (1999) 1521, hep-ex/9905013.}%
\nref\teexp{K.~W.~Edwards {\it et al.}, CLEO Collaboration,
 Phys. Rev. D55 (1997) 3919.}%
\nref\tmexp{S.~Ahmed {\it et al.}, CLEO Collaboration, hep-ex/9910060.}%
\refs{\meexp,\teexp,\tmexp},
\eqn\raddec{\eqalign{
B(\mu\ra e\gamma)\ \leq&\ 1.2\times10^{-11},\cr
B(\tau\ra e\gamma)\ \leq&\ 2.7\times10^{-6},\cr
B(\tau\ra\mu\gamma)\ \leq&\ 1.1\times10^{-6},\cr}}
and the bound on $\mu-e$ conversion
\ref\Wintz{P.~Wintz, Sindrum II Collaboration, in {\it Proceedings
of Lepton and Baryon Number Violation in Particle Physics,
Astrophysics and Cosmology}, eds. H.~V.~Klapdor-Kleingrothaus and
I.~V.~Krivosheina (Institute of Physics Publishing, Bristol and
Philadelphia, 1999), p.~534.},
\eqn\conversion{
{\sigma(\mu^- {\rm Ti} \ra e^- {\rm Ti}) \over
\sigma(\mu^- {\rm Ti} \ra {\rm capture})} < 6.1 \times 10^{-13}.}
In the future, all of these sensitivities are likely to improve.
Particularly promising are those involving muon decay and conversion
\ref\Okada{For a recent review, see Y.~Kuno and Y.~Okada,
hep-ph/9909265. See also contributions in {\it Proceedings of the
Workshop on Physics at the First Muon Collider and at the Front End of
a Muon Collider}, eds. S.~Geer and R.~Raja (American Institute of
Physics, Woodbury, New York, 1998).}: 
for example, a future experiment at PSI will be sensitive to $B(\mu\ra
e\gamma)$ at the $10^{-14}$ level
\ref\PSI{L.~M.~Barkov {\it et al.}, research proposal R--99--05,
``Search for $\mu^+ \ra e^+ \gamma$ at $10^{-14}$,'' 1999.}, and the
MECO collaboration has proposed an experiment to probe $\mu-e$
conversion down to $5\times 10^{-17}$, four orders of magnitude beyond
present sensitivities
\ref\Molzon{MECO and KOPIO Collaborations, ``A Proposal to the NSF 
to Construct the MECO and KOPIO Experiments,'' 1999,
http://meco.ps.uci.edu/RSVP.html; W.~Molzon, in {\it Proceedings of
Lepton and Baryon Number Violation in Particle Physics, Astrophysics
and Cosmology}, eds. H.~V.~Klapdor-Kleingrothaus and I.~V.~Krivosheina
(Institute of Physics Publishing, Bristol and Philadelphia, 1999),
p.~586.}.  In models where LFV is mediated dominantly by a photon, as
in the supersymmetric models discussed here, these rates are related
by~\ref\Czar{A.~Czarnecki, W.~J.~Marciano and K.~Melnikov,
hep-ph/9801218; see also Y.~G.~Kim, P.~Ko, J.~S.~Lee, and K.~Y.~Lee,
Phys.~Rev.~D59 (1999) 055018, hep-ph/9811211.}
\eqn\mueratio{
{\sigma(\mu^- {\rm Ti} \ra e^- {\rm Ti}) \over
\sigma(\mu^- {\rm Ti} \ra {\rm capture})} \approx 0.003\ B(\mu\ra
e\gamma).}
In the following, for brevity, we present predictions for decay rates
only, but it should be understood that the current bounds from and
prospects for $\mu \ra e \gamma$ and $\mu-e$ conversion are competitive.

The purpose of this work is to understand the implications of the
lepton flavor parameters of Eqs.~\ANpar, \SNpar\ and \chamas\ for the
lepton flavor changing processes of Eqs.~\raddec\ and \conversion. We
focus on the framework of supersymmetric Abelian horizontal
symmetries. (Similar issues have been investigated within different
supersymmetric flavor models in Refs.~\nref\LeTr{G.~K.~Leontaris 
and N.~D.~Tracas,
 Phys. Lett. B431 (1998) 90, hep-ph/9803320.}%
\nref\GLLV{M.~E.~Gomez, G.~K.~Leontaris, S.~Lola and J.~D.~Vergados,
 Phys. Rev. D59 (1999) 116009, hep-ph/9810291.}%
\nref\BDG{W.~Buchmuller, D.~Delepine and F.~Vissani,
 Phys. Lett. B459 (1999) 171, hep-ph/9904219.}%
\refs{\LeTr-\BDG}.) Our work is closely related to that of
Refs.~\nref\GNS{Y.~Grossman, Y.~Nir and Y.~Shadmi,
 JHEP 9810 (1998) 007, hep-ph/9808355.}%
\nref\NiSh{Y.~Nir and Y.~Shadmi,
 JHEP 9905 (1999) 023, hep-ph/9902293.}%
\refs{\GNS,\NiSh}\
where the various classes of models of Abelian flavor symmetries that
can accommodate \ANpar\ and \SNpar\ were presented.  Here we analyze
the consequences of these classes of models for charged LFV. We
will try to answer the following three questions:
\item{(i)} Are any of the relevant flavor models excluded by the upper
bounds on lepton flavor changing processes?
\item{(ii)} Are there generic (or, preferably, model-independent)
predictions for such processes in this framework that can be tested in
the future?
\item{(iii)} Is it possible to construct Abelian flavor models that
predict charged LFV far below even future sensitivities, and if so,
how complicated are these models?

Before we describe the details of our study, we emphasize that our
basic, underlying assumption is that Abelian flavor symmetries
determine the structure of both Yukawa couplings and supersymmetry
breaking parameters. (When we discuss models in which there are
additional ingredients that affect the hierarchy in the flavor
parameters, such as models without $R$-parity, we will state so
explicitly.) It could be, however, that Abelian flavor symmetries
determine the structure of the Yukawa couplings, but their effects on
the supersymmetry breaking parameters are screened. 
\nref\CEKLP{D. Choudhury, F. Eberlein, A. Konig, J. Louis
and S. Pokorski, Phys. Lett. B342 (1995) 180, hep-ph/9408275.}%
\nref\Eyal{G. Eyal, Phys. Lett. B461 (1999) 71, hep-ph/9903423.}%
This is the case, for example, if the slepton masses are dominated by
large, universal contributions from renormalization group
evolution. For squarks, the universal contribution from gaugino masses
could easily be dominant. For sleptons, however, the effects are much
weaker since here the renormalization is driven by $\alpha_2$ instead
of $\alpha_3$ \refs{\CEKLP,\Eyal}.  A more likely case that would lead
to slepton mass universality is supersymmetry breaking that is
mediated at some low energy in a flavor-blind way, as in
gauge-mediated supersymmetry breaking.  In such cases, our study does
not apply. In particular, when we state that various models in the
literature are excluded, we base our statements on the above
assumption. Most of these models are still viable models of neutrino
parameters if slepton masses are approximately universal.

\newsec{General Considerations}
\subsec{Supersymmetric contributions to charged LFV}

Supersymmetric models provide, in general, new sources of flavor
violation.  These are most commonly analyzed in the basis in which the
charged lepton mass matrix and the gaugino vertices are diagonal.  In
this basis, the slepton masses are not necessarily flavor-diagonal
and have the form
\eqn\slmass{ \tilde{\ell}^*_{M\, i}
(M^2_{\tilde{\ell}})^{MN}_{ij} \tilde{\ell}_{N\, j} =
\matrix{( \tilde{\ell}^*_{L\, i} & \tilde{\ell}^*_{R\, k})  }
\left(\matrix{M^2_{L\, ij} & A_{il} v_d \cr
         A_{jk} v_d & M^2_{R\, kl}} \right)
\left(\matrix{\tilde{\ell}_{L\, j} \cr \tilde{\ell}_{R\, l}} \right)
 \ ,} where $M,N = L,R$ label chirality, and $i,j,k,l = 1,2,3$ are
generational indices.  $M^2_L$ and $M^2_R$ are the supersymmetry
breaking slepton masses. The $A$ parameters enter in the trilinear
scalar couplings $A_{ij}\phi_d\tilde\ell_{Li}\tilde\ell_{Rj}^*$, where
$\phi_d$ is the down-type Higgs boson, and $v_d=\vev{\phi_d}$.  We
neglect small flavor-conserving terms involving $\tan\beta$, the ratio
of Higgs vacuum expectation values.

In this basis, charged LFV takes place through one or more slepton
mass insertion.  Each mass insertion brings with it a factor of
$\delta^{MN}_{ij} \equiv (M^2_{\tilde{\ell}})^{MN}_{ij} /
\tilde{m}^2$, where $\tilde{m}^2$ is the representative slepton mass
scale.  Physical processes therefore constrain
\eqn\deltaeff{(\delta^{MN}_{ij})_{\rm eff} \sim {\rm max} \left[
\delta^{MN}_{ij},\ \delta^{MP}_{ik} \delta^{PN}_{kj},\ \ldots,\
( i \leftrightarrow j) \right] \ . }
For example,
\eqn\exeff{(\delta^{LR}_{12})_{\rm eff}
\sim{\rm max}\left[ A_{12} v_d/\tilde m^2,\
M^2_{L1k}A_{k2}v_d/\tilde m^4,\ A_{1k}v_dM^2_{Rk2}/\tilde m^4,\
\ldots,\ ( 1 \leftrightarrow 2) \right].}
Note that contributions with two or more insertions may be less
suppressed than those with only one.

In models with horizontal symmetries, Yukawa and supersymmetry
breaking parameters are ambiguous up to ${\cal O}(1)$ factors.  It is
therefore sufficient to obtain order of magnitude estimates for the
supersymmetric contributions to radiative lepton decay.  These have
been analyzed in \ref\GGMS{F.~Gabbiani, E.~Gabrielli, A.~Masiero and
L.~Silvestrini, Nucl. Phys. B477 (1996) 321, hep-ph/9604387.}.
Normalizing these results to the current bounds, we find
\eqn\LFVKij{\eqalign{{
B(\mu\ra e\gamma)\over1.2\times10^{-11}}\ \sim&\ {\rm max}\left[
\left({(\delta^{LL}_{12})_{\rm eff}\over2.0\times10^{-3}}\right)^2,
\left({(\delta^{LR}_{12})_{\rm eff}\over6.9\times10^{-7}}\right)^2\right]
\left({100\ {\rm GeV}\over\tilde m}\right)^4,\cr
{B(\tau\ra e\gamma)\over2.7\times10^{-6}}\ \sim&\ {\rm max}\left[
\left({(\delta^{LL}_{13})_{\rm eff}\over2.2}\right)^2,
\left({(\delta^{LR}_{13})_{\rm eff}\over1.3\times10^{-2}}\right)^2\right]
\left({100\ {\rm GeV}\over\tilde m}\right)^4,\cr
{B(\tau\ra \mu\gamma)\over1.1\times10^{-6}}\ \sim&\ {\rm max}\left[
\left({(\delta^{LL}_{23})_{\rm eff}\over1.4}\right)^2,
\left({(\delta^{LR}_{23})_{\rm eff}\over8.3\times10^{-3}}\right)^2\right]
\left({100\ {\rm GeV}\over\tilde m}\right)^4.\cr}}
Here, the lightest neutralino is assumed to be photino-like, and so
bounds on $(\delta^{LL}_{ij})_{\rm eff}$ apply also to
$(\delta^{RR}_{ij})_{\rm eff}$.  We have set $m^2_{\tilde\gamma}/
\tilde m^2=0.3$.  The bounds are fairly insensitive to this ratio; for
example, even for $m^2_{\tilde\gamma}/ \tilde m^2=1$, the bounds on
$\delta^{LL}$ and $\delta^{LR}$ are only weakened by factors of $\sim
2$ and 1.2, respectively~\refs{\GGMS}.

Finally, we note that in the physical basis with diagonal charged
lepton and slepton masses, flavor violation appears in the gaugino
vertices $K^{MN}_{ij} \tilde{\gamma} \ell_{M\, i} \tilde{\ell}^*_{N\,
j}$.  (Of course, left- and right-handed sleptons mix; in our notation
here, the slepton mass eigenstates are labeled by their dominant
chirality.)  The mass insertion parameters are related to these mixing
angles by $(\delta^{MN}_{ij})_{\rm eff} \sim {\rm max} \left[
K^{MP}_{ik} K^{NP}_{jk},\ (i \leftrightarrow j)
\right]$.

\subsec{Models of Abelian flavor symmetries}

We are interested in finding the relevant $(\delta^{MN}_{ij})_{\rm
eff}$ parameters in the framework of approximate Abelian flavor
symmetries. We have in mind theories with a spontaneously broken
horizontal symmetry of one of the following three types: $(i)$ An
anomalous $U(1)$ symmetry where the anomaly is cancelled by the
Green-Schwarz mechanism; $(ii)$ A discrete $Z_n$ symmetry; $(iii)$ A
non-anomalous $U(1)$. The details of these full high energy theories
are not important: for our purposes, it is sufficient to consider low
energy effective theories with a horizontal symmetry that is
explicitly broken by a small parameter. The three types of such models
that we consider are, however, motivated by the high energy theories
described above. We define the models by the selection rules that
apply to the low energy effective theory:

$(i)$ {\it A $U(1)$ symmetry broken by a single breaking parameter.}
We denote the breaking parameter by $\lambda$ and assign to it a
horizontal charge $-1$.  Then the following selection rules apply:
\item{a.} Terms in the superpotential that carry an integer $U(1)$ charge
$n\geq0$ are suppressed by $\lambda^n$.  Terms with $n<0$ vanish by
holomorphy.
\item{b.} Terms in the K\"ahler potential that carry an integer $U(1)$
charge $n$ are suppressed by $\lambda^{|n|}$.

$(ii)$ {\it A $Z_m$ symmetry broken by a single breaking parameter.}
We denote the breaking parameter by $\lambda$ and assign to it a
horizontal charge $-1$.  Then the following selection rules apply:
\item{a.} Terms in the superpotential that carry an integer $Z_m$
charge $n$ are suppressed by $\lambda^{n\ ({\rm mod}\ m)}$.
\item{b.} Terms in the K\"ahler potential that carry an integer $Z_m$
charge $n$ are suppressed by $\lambda^{{\rm min}[|n|,n\ ({\rm mod}\
m)]}$.

$(iii)$ {\it A $U(1)$ symmetry broken by two breaking parameters.} We
denote the breaking parameters by $\lambda$ and $\bar\lambda$. They
have equal magnitude, $\lambda=\bar\lambda$, and carry opposite
horizontal charge, $+1$ and $-1$, respectively. Then the following
selection rules apply:
\item{a.} Terms in the superpotential and in the K\"ahler potential
that carry an integer $U(1)$ charge $n$ are suppressed by
$\lambda^{|n|}$.

\noindent In all cases, terms in both the superpotential and K\"ahler
potential with non-integer horizontal charge vanish.

To be specific, we set $\lambda\sim0.2$ and require that our models
are consistent with \ANpar, \SNpar\ and \chamas, namely that they give
the following parametric suppression for the lepton flavor parameters:
\eqn\parmix{V_{23}^\ell\sim1,\ \ \ V_{13}^\ell\lsim\lambda,\ \ \
V_{12}^\ell\sim\cases{1&MSW(LA), VO\cr \lambda^2&MSW(SA)\cr},}
\eqn\parmnu{{\Delta m^2_{12}\over \Delta m^2_{23}}\sim\cases{
\lambda^2-\lambda^4&MSW\cr \lambda^8-\lambda^{12}&VO\cr},}
\eqn\parmch{m_\tau/\vev{\phi_d}\sim\lambda^3-1,\ \ \
m_\mu/m_\tau\sim\lambda^2,\ \ \ m_e/m_\mu\sim\lambda^3.}
In \parmnu, we allow a large range for the VO option, since when
observables depend on a very high power of $\lambda$, the sensitivity
to the precise value of the breaking parameter is enhanced. In \parmch,
the range for the Yukawa coupling of the tau corresponds to values
of $\tan\beta$ between 1 and $m_t/m_b$.

\subsec{A naive estimate}

Before we analyze specific classes of models of Abelian flavor
symmetries, let us introduce a naive estimate of the
$(\delta^{MN}_{ij})_{\rm eff}$ parameters in this framework. By `naive
estimate' we mean that we make an order of magnitude estimate in
models with the following features:
\item{a.} The horizontal symmetry is a single $U(1)$;
\item{b.} The symmetry is broken by a single parameter;
\item{c.} Holomorphic zeros play no role;
\item{d.} Singlet neutrinos play no role.

\noindent  We emphasize that such `naive models' cannot accommodate
the neutrino parameters of Eqs.~\ANpar\ and \SNpar. In particular, as
alluded to previously and as will become clear, such models cannot
explain large mixings between states with hierarchically different
masses.  Therefore, we should not expect that our naive predictions
necessarily hold in all viable models.  However, naive models and
viable models often share several important features, and so it is a
worthwhile exercise to consider first the simpler case of naive
models.

In this framework, we have the following order of magnitude estimates
in the interaction basis, where the horizontal charges are
well-defined:
\eqn\naivemat{\eqalign{
(M_\ell)_{ij}\ \sim&\ v_d\lambda^{H(L_i)+H(\bar\ell_j)+H(\phi_d)},\cr
(M_\nu)_{ij}\ \sim&\ {v_u^2\over
M}\lambda^{H(L_i)+H(L_j)+2H(\phi_u)},\cr (M^2_L)_{ij}\ \sim&\ \tilde
m^2\lambda^{|H(L_i)-H(L_j)|},\cr (M^2_R)_{ij}\ \sim&\ \tilde
m^2\lambda^{|H(\bar\ell_i)-H(\bar\ell_j)|},\cr A_{ij}\ \sim&\ \tilde
m\lambda^{H(L_i)+H(\bar\ell_j)+H(\phi_d)}.\cr}} Here $L_i$ are lepton
doublets and $\bar\ell_i$ are charged lepton singlets.  The Higgs
vacuum expectation values are denoted by $v_u$ and $v_d$, and $M$ is
some large mass scale.  Hypercharge and Peccei-Quinn U(1) symmetries
may be used to set the charges of both Higgs bosons to zero, and we
indeed do so in all the explicit models described below.\foot{Note,
however, that the Peccei-Quinn symmetry is not a symmetry of the full
theory. It is broken explicitly by the $\mu$ term. It is only an
accidental symmetry of the Yukawa sector, of the $A$-couplings and of
the slepton mass-squared terms.  Shifts by the Peccei-Quinn charge do
not affect these sectors and we can use them for our purposes.}

{}From \naivemat, one can deduce the order of magnitude of the
physical parameters. In particular, for the mixing angles in the
$W^\pm$ couplings to neutrinos and charged leptons $V^\ell_{ij}$ (that
is, the MNS matrix
\ref\MNS{Z.~Maki, M.~Nakagawa and S.~Sakata,
 Prog. Theo. Phys. 28 (1962) 247.})
and for the mixing angles in the photino $\tilde\gamma$ couplings to
charged leptons and charged sleptons, $K^{MN}_{ij}$, we have
\eqn\naivemix{\eqalign{
V^\ell_{ij}\ \sim&\ \lambda^{|H(L_i)-H(L_j)|},\cr
K^{LL}_{ij}\ \sim&\ \lambda^{|H(L_i)-H(L_j)|},\cr
K^{RR}_{ij}\ \sim&\ \lambda^{|H(\bar\ell_i)-H(\bar\ell_j)|},\cr
K^{LR}_{ij}\ \sim&\ (v_d/\tilde m)
\lambda^{H(L_i)+H(\bar\ell_j)+H(\phi_d)},\cr}}
while for the fermion masses we have
\eqn\naivemas{\eqalign{
m(\ell^\pm_i)\ \sim&\ v_d \lambda^{H(L_i) + H(\bar\ell_i) + 
H(\phi_d)} ,\cr
m(\nu_i)\ \sim&\ {v_u^2 \over M} \lambda^{2H(L_i)+2H(\phi_u)}.\cr}}

Equations \naivemix\ and \naivemas\ demonstrate in a clear way how the
fermion flavor parameters (masses and mixing) are related to the
supersymmetric flavor violation. Explicitly, we get:
\eqn\susysm{\eqalign{
K^{LL}_{ij}\ \sim&\ V^\ell_{ij},\cr
K^{RR}_{ij}\ \sim&\ {m(\ell^\pm_i)\over m(\ell^\pm_j)V^\ell_{ij}},\cr
K^{LR}_{ij}\ \sim&\ {m(\ell^\pm_j)V^\ell_{ij}\over\tilde m}, \
{\rm where}\ H(L_i) > H(L_j)\ {\rm and}\ H(\bar\ell_i) > 
H(\bar\ell_j)\cr
K^{LR}_{ji}\ \sim&\ {m(\ell^\pm_i)\over V^\ell_{ij}\tilde m}, \quad
{\rm where}\ H(L_i) > H(L_j)\ {\rm and}\ H(\bar\ell_i) > 
H(\bar\ell_j).\cr}}

We can use Eqs.~\ANpar, \SNpar\ and \chamas\ to make naive predictions
for the relevant supersymmetric mixing angles.  For the $\tau\ra\mu$
transitions, we have:
\eqn\naitm{\eqalign{
(\delta^{LL}_{23})_{\rm eff}\ \sim&\ V^\ell_{23}\sim1,\cr
(\delta^{LR}_{23})_{\rm eff}\ \sim&\ {m_\tau V^\ell_{23}\over\tilde m}
\sim0.02 \left({100\ {\rm GeV}\over\tilde m}\right).\cr}}
For the $\mu\ra e$ transitions, we have for both MSW(LA) and VO:
\eqn\naimeL{\eqalign{
(\delta^{LL}_{12})_{\rm eff}\ \sim&\ V^\ell_{12}\sim1,\cr
(\delta^{LR}_{12})_{\rm eff}\ \sim&\ {m_\mu V^\ell_{12}\over\tilde m}
\sim10^{-3}
\left({100\ {\rm GeV}\over\tilde m}\right),\cr}}
and for MSW(SA):
\eqn\naimeS{\eqalign{
(\delta^{LL}_{12})_{\rm eff}\ \sim&\ V^\ell_{12}\sim0.04,\cr
(\delta^{LR}_{12})_{\rm eff}\ \sim&\ {m_e\over V^\ell_{12}\tilde m}
\sim10^{-4}
\left({100\ {\rm GeV}\over\tilde m}\right).\cr}}
For the $\tau\ra e$ transitions, we have for both MSW(LA) and VO:
\eqn\naiteL{\eqalign{
(\delta^{LL}_{13})_{\rm eff}\ \sim&\ V^\ell_{12}V^\ell_{23}\sim1,\cr
(\delta^{LR}_{13})_{\rm eff}\ \sim&\ {m_\tau V^\ell_{13}\over\tilde m}
\sim0.02 \left({100\ {\rm GeV}\over\tilde m}\right),\cr}}
and for MSW(SA):
\eqn\naiteS{\eqalign{
(\delta^{LL}_{13})_{\rm eff}\ \sim&\ V^\ell_{12}V^\ell_{23}\sim0.04,\cr
(\delta^{LR}_{13})_{\rm eff}\ \sim&\ {m_\tau V^\ell_{13}\over\tilde m}
\sim10^{-3} \left({100\ {\rm GeV}\over\tilde m}\right).\cr}}

We emphasize that the parameters of order one could be accidentally
large or small, leading to an incorrect `translation' of the
experimental numbers to powers of $\lambda$ in Eqs.~\parmix$-$\parmch,
or to deviations of the $(\delta_{ij}^{MN})_{\rm eff}$ from the
numerical estimates in Eqs.~\naitm$-$\naiteS. These ambiguities
constitute a limitation to the predictive power of this framework. We
avoid, however, part of this ambiguity by presenting our estimates of
the $(\delta_{ij}^{MN})_{\rm eff}$ parameters in
Eqs.~\naitm$-$\naiteS\ in terms of mixing angles and mass ratios
rather than directly in powers of $\lambda$. Our point is that the
parametric suppression of the slepton flavor parameters is the same as
that of the corresponding combinations of lepton flavor parameters.
This statement is independent of the parameters of order one.

A comparison of the naive estimates of Eqs.~\naitm--\naiteS\ with
Eq.~\LFVKij\ leads to the following conclusions:
\item{(i)} The MSW(LA) and VO solutions of the solar neutrino problem
cannot be accommodated.
\item{(ii)} The MSW(SA) can be accommodated if the sleptons are heavier
than ${\cal O}(500\ {\rm GeV})$. The rate of the $\mu\ra e\gamma$ decay
should be close to the present bound.
\item{(iii)} The rate of the $\tau\ra\mu\gamma$ decay should be not far
below the present bound. (It is within one order of magnitude
of the present bound if $\tilde m\lsim200\ GeV$, but falls 
like $1/\tilde m^6$ up to values of $\tilde m\sim350\ GeV$ and
like $1/\tilde m^4$ for higher values.)

As noted above, `naive models,' which obey the conditions a-d
specified above, cannot accommodate the neutrino parameters of Eqs.
\ANpar\ and \SNpar. In particular, the first relation of
Eq.~\naivemix\ and the last of Eq.~\naivemas\ require highly mixed
neutrinos to have similar masses. Note, however, that
Eqs.~\naitm$-$\naiteS\ relate the $(\delta_{ij}^{MN})_{\rm eff}$
mixing angles to the MNS mixing angles $V_{ij}^\ell$ and charged
lepton masses, but not to the neutrino masses. Many of the viable
extensions of the naive models modify Eq.~\naivemas\ for the neutrino
mass ratios but not Eq.~\naivemix\ for the mixing
angles. Consequently, the naive estimates remain valid in a large
class of models. For such models, our analysis in this section gives
the following important lessons:
\item{1.} Models of MSW(LA) or of VO where the naive predictions for
$(\delta_{12}^{LL})_{\rm eff}$ and $(\delta_{12}^{LR})_{\rm eff}$ hold
are excluded.
\item{2.} Both $\mu\ra e\gamma$ and $\tau\ra\mu\gamma$ decays provide
interesting probes of Abelian flavor symmetries. The $\tau\ra e\gamma$
decay is, at present, less sensitive to this type of new physics.

\subsec{Pseudo-Dirac neutrinos}
The AN data imply a large, maybe maximal mixing in the 23 subspace.
The MSW(LA) and VO solutions of the SN problem require large mixing
in the 12 space. It is an interesting possibility then that two
of the neutrinos form a pseudo-Dirac neutrino, which would yield close
to maximal mixing. This scenario becomes even more attractive in the
framework of Abelian horizontal symmetries, because the symmetry could
easily force two neutrinos into a pseudo-Dirac structure
\ref\BLPR{P.~Binetruy, S.~Lavignac, S.~Petcov and P.~Ramond,
 Nucl. Phys. B496 (1997) 3, hep-ph/9610481.}.  Take, for example, a
horizontal $U(1)$ symmetry where two lepton doublets carry opposite
charges (and $H(\phi_u)=0$). Then, the corresponding off-diagonal
terms in the Majorana mass matrix carry no $U(1)$ charge and are
therefore unsuppressed by the horizontal symmetry. The diagonal terms,
on the other hand, carry horizontal charges, and are either suppressed
or forbidden. To be specific, take $H(L_2)=-1$ and $H(L_1)=+1$. Then,
in the 12 subspace,
\eqn\psDi{M_{\nu}\sim{v_u^2\over M}\pmatrix{\lambda^2&1\cr1&0\cr},}
yielding $\sin^22\theta_{12}^\nu\simeq1$.

In Ref.~\NiSh\ it was argued, however, that in Abelian flavor models
that satisfy both \ANpar\ and \SNpar, the pseudo-Dirac structure
cannot apply to the 23 subspace. It can only be relevant then to the
12 subspace, corresponding to either the MSW(LA) or the VO solution of
the SN problem.

The interesting point about the case where $\nu_e$ and $\nu_\mu$ form
a pseudo-Dirac neutrino is that the large mixing, $V_{12}^\ell\sim1$,
comes from the neutrino sector. It is therefore not necessary that the
charged lepton sector induce large 12 mixing. In fact,
$\sin\theta^\ell_{12}\ll1$ is unavoidable in models where a horizontal
$U(1)$ is broken by a single parameter and it is generic (though not
unavoidable) in models where the symmetry is broken by two small
parameters of opposite signs.  The naive predictions of \naimeL\ are
therefore avoided.

We learn that if, in the future, measurements of SN make a convincing
case for a (close to) maximal mixing but $\mu\ra e\gamma$ is not
observed, then a pseudo-Dirac structure for the corresponding
neutrinos induced by an Abelian flavor symmetry can provide a very
attractive explanation for this situation.

This statement is particularly relevant to the case of VO.  In the
case of MSW(LA), a truly maximal mixing is disfavored (see, {\it
e.g.}, the discussion in
Ref.~\ref\BHSSW{R.~Barbieri, L.~J.~Hall, D.~Smith, A.~Strumia and
N.~Weiner, JHEP 9812 (1998) 017, hep-ph/9807235.}).
If $\nu_e$ and $\nu_\mu$ form a pseudo-Dirac neutrino, a sufficient
deviation from maximal mixing can be induced by
$\sin\theta_{12}^\ell$:
\eqn\MSWpd{\sin\theta_{12}^\nu=\sqrt2/2,\ \
\sin\theta_{12}^\ell\gsim0.3\
\ \ \Longrightarrow\ \ \ \sin^22\theta_{12}=1-{\cal O}(0.1).}
However, in such a scenario, we have
\eqn\naipD{\eqalign{
(\delta^{LL}_{12})_{\rm eff}\ \sim&\ 0.3,\cr (\delta^{LR}_{12})_{\rm
eff}\ \sim&\ 3\times10^{-4}
\left({100\ {\rm GeV}\over\tilde m}\right).\cr}}
Comparing \naipD\ and \LFVKij, we conclude that viable MSW(LA) models
with pseudo-Dirac structure in the 12 subspace require $\tilde
m\gsim1\ {\rm TeV}$ and are, therefore, disfavored.

\newsec{Specific Models}

The naive models discussed above cannot accommodate the neutrino
parameters of~\ANpar\ and \SNpar.  In this section, we survey specific
supersymmetric models with Abelian flavor symmetries that have been
constructed to accommodate these parameters, and find their
predictions for the lepton flavor violating decays \raddec.

\subsec{Accidental mass hierarchy}

As mentioned above, the main problem in accommodating the neutrino
parameters is to have simultaneously a large mixing,
$V^\ell_{23}\sim1$, and a large hierarchy, $m_2/m_3\lsim0.1$. In
the case of MSW solutions to the solar neutrino problem, however, the
hierarchy is close to $0.1$ and could be simply accidental. By
`accidental' we mean that the hierarchy does not result from
suppression by a small symmetry breaking parameter. Instead, it is the
result of an accidental cancellation between ${\cal O}(1)$
coefficients, {\it e.g.}, $ac-b^2={\cal O}(0.1)$ with $a,b,c= {\cal
O}(1)$. Such models generically allow a situation where $L_2$ and
$L_3$ carry the same horizontal charge. In this case, we have
\eqn\acctwth{(M_\ell)_{23}/(M_\ell)_{33}\sim1\ \ \Longrightarrow\ \
(\delta^{LL}_{23})_{\rm eff}\sim1,\ \ (\delta^{LR}_{23})_{\rm
eff}\sim0.02\left({100\ {\rm GeV}\over\tilde m}\right),} as in the
naive prediction of Eq.~\naitm. Therefore, a generic (though not an
unavoidable) prediction of this class of models is that, if charged
slepton masses are not much higher than 100 GeV, $B(\tau\ra\mu\gamma)$
should be close to the upper bound.  As concerns $B(\mu\ra e\gamma)$,
there is no generic prediction here.

Abelian flavor models of this type were constructed in
Refs.~\nref\ILR{N.~Irges, S.~Lavignac and P.~Ramond,
 Phys. Rev. D58 (1998) 035003, hep-ph/9802334.}%
\nref\EIR{J.~K.~Elwood, N.~Irges and P.~Ramond,
 Phys. Rev. Lett. 81 (1998) 5064, hep-ph/9807228.}%
\refs{\ILR,\EIR}.
Equation \acctwth\ holds, indeed, in these models.  The explicit
models are constructed to accommodate the MSW(SA) solution of the SN
problem. They actually give a rather small value for the relevant
mixing angle, that is,
\eqn\acctwon{ (M_\ell)_{12}/(M_\ell)_{22}\sim\lambda^3\ \
\Longrightarrow\ \
(\delta^{LL}_{12})_{\rm eff}\sim8\times10^{-3},\ \
(\delta^{LR}_{12})_{\rm eff}\sim6\times10^{-4}
\left({100\ {\rm GeV}\over\tilde m}\right).}
(The naive estimates \naimeS\ hold in these models, except that the
value of $V_{12}$ is smaller than what is implied by \SNpar. The
appropriate mixing in the charged current interactions is accidentally
enhanced by about one order of magnitude.) These models are then
viable, provided that slepton masses are rather heavy, $\tilde
m\gsim500\ {\rm GeV}$.

\subsec{Neutrino masses from different sources}

Different neutrino masses could come from different sources, so that
the mass hierarchy is determined not only by the horizontal
symmetry. In such a framework, neutrinos with the same horizontal
charge, and therefore with large mixing, may nevertheless have their
masses hierarchically separated. The problem of accommodating \ANpar\
and \SNpar\ is then solved.  It is then generic (though, again, not
unavoidable) in these models that the horizontal charges of $L_2$ and
$L_3$ are equal, leading to \acctwth\ and, consequently, to
$B(\tau\ra\mu\gamma)$ close to the bound.

A framework where this is the case is that of supersymmetry without
$R$-parity. The Abelian horizontal symmetry could replace $R$-parity
in suppressing dangerous lepton-number violating couplings
\ref\Bank{T.~Banks, Y.~Grossman, E.~Nardi and Y.~Nir,
 Phys. Rev. D52 (1995) 5319, hep-ph/9505248.}.  If the $B$- and
$\mu$-terms are not aligned, one neutrino acquires its mass at tree
level by mixing with neutralinos, while the other two acquire masses
at the loop level.  Explicit flavor models of this type were
constructed, for example, in Refs.~\nref\BGNN{F.~M.~Borzumati,
Y.~Grossman, E.~Nardi and Y.~Nir, Phys. Lett. B384 (1996) 123,
hep-ph/9606251.}%
\nref\CCH{K.~Choi, E.~J.~Chun and K.~Hwang,
 Phys. Rev. D60 (1999) 031301, hep-ph/9811363.}%
\nref\HVFK{O.~Haug, J.~D.~Vergados, A.~Faessler and S.~Kovalenko,
  hep-ph/9909318.}%
\refs{\BGNN,\CCH,\HVFK}.
Equation \acctwth\ holds in these models.  The explicit models are
constructed to accommodate the MSW(SA) solution of the SN problem and
Eq.~\naimeS\ holds. (Ref.~\CCH\ assumes that supersymmetry breaking is
gauge-mediated, in which case slepton masses are degenerate and the
radiative lepton decays are highly suppressed.)

\subsec{See-saw enhancement}

A neutrino mass could be enhanced beyond the naive expectation by the
see-saw mechanism. Singlet neutrinos with masses below the scale of
horizontal symmetry breaking could induce such an enhancement \GNS.
In such a framework it is, again, possible that $L_2$ and $L_3$ carry
the same horizontal charges. Consequently, the generic prediction is
that of Eq.~\acctwth.

The idea of see-saw enhancement was presented in Ref.~\BHSSW.  The
horizontal symmetry is simply $L_e-L_\mu-L_\tau$. Under this symmetry,
$H(L_2)=H(L_3)=-1$ and, consequently, \naitm\ holds.

Interestingly, the symmetry forces a pseudo-Dirac neutrino in the 12
subspace.  The large $V_{12}^\ell$ comes from the neutrino sector. The
12 mixing in the charged lepton sector is parametrically suppressed
but, when fitted to MSW(LA) parameters, it is found that numerically
the suppression is mild. Equation \naipD\ holds, so that the model is
viable only for very high slepton masses, $\tilde m\gsim1\ {\rm TeV}$.

Another model in Ref.~\BHSSW, within the class of see-saw enhancement,
is constructed to fit the MSW(SA) parameters. The naive predictions
\naitm\ and \naimeS\ for, respectively, $(\delta^{MN}_{23})_{\rm eff}$
and $(\delta^{MN}_{12})_{\rm eff}$, are valid. This model requires
then that $\tilde m\gsim500\ {\rm GeV}$.

A framework where a horizontal $U(1)$ is combined with an $SU(5)$
grand unified theory is presented in
Refs.~\nref\AlFe{G.~Altarelli and F.~Feruglio,
 Phys. Lett. B451 (1999) 388, hep-ph/9812475; hep-ph/9905536.}%
\nref\ShTa{Q.~Shafi and T.~Tavartkiladze, Phys. Lett. B451 (1999) 129,
 hep-ph/9901243.}%
\refs{\AlFe,\ShTa}.
Here, $H(L_2)=H(L_3)$ and the hierarchy of masses is induced by
see-saw enhancement. A large 23 mixing, as in \naitm, is predicted.
As concerns the 12 mixing, the largest effect comes from
$\delta^{RR}_{12}\sim\lambda$. These specific models are then also
only viable for very high slepton masses, $\tilde m\gsim1\ {\rm TeV}$.

A framework where a horizontal $U(1)$ is broken into an exact lepton
parity was presented in Ref.~\ref\EyNi{G.~Eyal and Y.~Nir, JHEP 9906
(1999) 024, hep-ph/9904473.}.  For the neutrino spectrum, the
mechanism of see-saw enhancement is in operation. In the specific
model presented in \EyNi, $H(L_2)=H(L_3)$ and \naitm\ holds. While the
model is constructed to fit the MSW(SA) parameters, it has
$H(\bar\ell_1)=H(\bar\ell_2)$, leading to a surprisingly large 12
mixing, $\delta^{RR}_{12}\sim1$. This specific model is then excluded.

\subsec{Holomorphic zeros}

Holomorphy could induce a strong suppression of a neutrino mass ratio,
compared to the naive estimate. This mechanism was proposed and viable
models were constructed in Ref.~\GNS. In the models of Ref.~\GNS,
singlet neutrinos play no role. Then, the mass matrix for the active
neutrinos in the 23 subspace is near-diagonal, and the large 23 mixing
must arise from the charged lepton sector.  Eq.~\naitm\ then holds
independently of the details of the model, and $\tau \ra \mu \gamma$
may be near its current bound.

As concerns the SN problem, the structure of the neutrino mass matrix
allows only a pseudo-Dirac structure in the 12 subspace. The
predictions of \naipD\ hold in the MSW(LA) case, requiring $\tilde
m\gsim1\ {\rm TeV}$ for the model to be viable. In the VO case,
$(\delta^{MN}_{12})_{\rm eff}$ could be very small and \naimeS\ does
not hold. In the specific example of Ref.~\GNS,
$(\delta^{LL}_{12})_{\rm eff}\sim\lambda^4$ and $B(\mu\ra e\gamma)$ is
close to the bound only if the sleptons are light, that is, $\tilde
m\sim100\ {\rm GeV}$.

\subsec{Discrete symmetries}

A discrete horizontal symmetry can lead to a large mixing
simultaneously with large hierarchy by modifying the predictions of a
continuous symmetry in one of the following three ways \GNS:
\item{1.} Mass enhancement: $(M_\nu)_{33}$ is enhanced;
\item{2.} Mixing enhancement: $(M_{\ell^\pm})_{23}$ is enhanced;
\item{3.} See-saw suppression: singlet neutrino masses are enhanced,
thus suppressing light neutrino masses.

In the models of Ref.~\GNS, singlet neutrinos play no role.  In the
first two cases, the discrete symmetry induces
$\sin\theta_{23}^\nu\ll1$ and therefore the large AN mixing must arise
in the charged lepton sector.  Consequently, \naitm\ necessarily
holds.  The third scenario is operative even in the case that
$H(L_2)=H(L_3)$.  The generic (though, in this case, not unavoidable)
prediction is then again that \naitm\ holds.

As concerns the 12 subspace, there is no generic prediction. The
explicit models constructed in Ref.~\GNS\ include an MSW(SA) model
where \naimeS\ holds and a VO model where \naimeL\ does not hold
($(\delta^{LL}_{12})_{\rm eff}\ll1$).

\subsec{Two breaking parameters}

If a horizontal $U(1)$ symmetry is broken by two small parameters of
equal magnitude and opposite charge
\ref\DLLRS{H.~ Dreiner, G.~K.~Leontaris, S.~Lola, G.~G.~Ross amd
C.~Scheich, Nucl. Phys. B436 (1995) 461, hep-ph/9409369.},
then a large mixing angle could arise also for $H(L_2)\neq H(L_3)$,
leading to hierarchical neutrino masses. Models of this type that
accommodate \ANpar\ and \SNpar\ have been constructed in Ref.~\NiSh.
In the models of \NiSh, the large mixing in the 23 subspace is
achieved by
\eqn\twbr{|H(L_2)+H(\bar\ell_3)|=|H(L_3)+H(\bar\ell_3)|.}
Then \acctwth\ holds with the resulting predictions for
$B(\tau\ra\mu\gamma)$.

As concerns the 12 subspace, there is again no generic prediction.  In
one model of Ref.~\NiSh, the VO solution is accommodated using the
same mechanism to induce the large 12 mixing as the 23 mixing, that
is, $|H(L_1)+H(\bar\ell_2)|=|H(L_2)+H(\bar\ell_2)|$. Consequently,
\eqn\twotwon{ (M_\ell)_{12}/(M_\ell)_{22}\sim1\ \ \Longrightarrow\ \
(\delta^{LL}_{12})_{\rm eff}\sim1,\ \
(\delta^{LR}_{12})_{\rm eff}\sim10^{-3}.}
In other words, the naive estimate of Eq.~\naimeL\ holds and the model
is excluded.

In another model of Ref.~\NiSh, the VO parameters are related to a
pseudo-Dirac structure in the 12 subspace. The charged lepton mass
matrix is near-diagonal in the 12 subspace and, consequently,
$(\delta^{MN}_{12})_{\rm eff}$ is negligibly small.

To summarize this section: in all the models that have been proposed
in the literature, the charged lepton sector has a large 23
mixing. Our naive estimate \naitm\ holds. If $\tilde m\sim100\ {\rm
GeV}$, then $B(\tau\ra\mu\gamma)$ is close to the upper bound.

On the other hand, there is a large variety of predictions concerning
$B(\mu\ra e\gamma)$. Some models are excluded because they predict
this rate to be above the present bound by many orders of magnitude.
Others are viable only if $\tilde m\gsim500\ {\rm GeV}$ and predict
that the rate of radiative muon decay is close to the bound. Finally,
there are models in which the charged lepton sector has a negligible
12 mixing, predicting $B(\mu\ra e\gamma)$ well below the present
bound.

\newsec{Avoiding $(\delta^{LL}_{23})_{\rm eff}\sim1$}

In our survey of the literature in the previous section, we have only
encountered models where the naive predictions \naitm\ hold and,
therefore, if slepton masses are of order 100 GeV, the rate of
$\tau\ra\mu\gamma$ is predicted to be close to the bound. This is not
an accidental result:
\item{(i)} In many models, $H(L_2)=H(L_3)$. Then, large
$(\delta^{LL}_{23})_{\rm eff}$ is unavoidable.
\item{(ii)} In many models, the hierarchy in the neutrino sector is
closely related to a near-diagonal structure of the neutrino mass
matrix in the 23 subspace. Then the charged lepton sector must account
for the large 23 mixing and $(\delta^{LL}_{23})_{\rm eff}\sim1$ is
unavoidable.

The second argument holds, however, only in models where singlet
neutrinos play no role. More precisely, it is valid in models where
the AN parameters are determined by the horizontal charges of $L_2$,
$L_3$, $\bar\ell_2$, $\bar\ell_3$ and the Higgs doublets only. (In
Ref.~\NiSh, these models are called (2,0) models, where the first
integer refers to the number of active neutrinos, and the second to
the number of sterile.) In such models, the selection rules apply
directly to the Majorana mass matrix of the active neutrinos, and it
is easy to see that the neutrino mass matrix cannot give both large
mixing and hierarchically separated masses: Large 23 mixing would
require $(M_\nu)_{23}\sim (M_\nu)_{33}$. Since an Abelian symmetry
cannot relate the coefficients of order one, this then implies that in
the 23 subspace $\det M_\nu\sim[(M_\nu)_{33}]^2$, and the two mass
eigenvalues are of the same order of magnitude.

As we will see below, however, if singlet neutrinos play a role in
determining the light neutrino masses and mixings, it is possible to
obtain both large mixing and mass hierarchies from the neutrino matrix
alone.  In this case, there need not be large mixings in the charged
lepton sector.  We will first argue that the addition of singlet
neutrinos by itself cannot lead to a situation where
$V^\ell_{23}\sim1$ and $(\delta^{LL}_{23})_{\rm eff}\ll1$; an
additional special ingredient, such as holomorphic zeros, is
required. Then we give three examples of such models, each related to
a different framework presented in the previous section.  These
illustrative examples are (2,2) models.  In the final subsection, we
show that it is also possible to suppress $(\delta^{LL}_{23})_{\rm
eff}$ in three generation models.  We first give a (3,3) example. We
then present a viable (3,0) model, that is, a model where singlet
neutrinos play no role but the horizontal charge of $L_1$ affects the
masses and/or mixing of $\nu_2$ and $\nu_3$.

\subsec{Naive (3,3) models}

We now argue that if we only modify our naive models, defined in
section 2.3, by the addition of singlet neutrinos, then we cannot
achieve $V^\ell_{23}\sim1$ and $(\delta^{LL}_{23})_{\rm eff}\ll1$. To
be specific, we consider a (3,3) model with the following features:
\item{a.} The horizontal symmetry is a single $U(1)$;
\item{b.} The symmetry is broken by a single parameter $\lambda(-1)$;
\item{c.} There are three active and three singlet neutrinos.
\item{d.} There are no holomorphic zeros in
either $M_\nu^{\rm Dir}$ or $M_{\nu_s}^{\rm Maj}$.

We denote the Majorana mass matrix for the {\it light} neutrinos
by $M_\nu$. It is given by
\eqn\Mlight{M_\nu=M_\nu^{\rm Dir}(M_{\nu_s}^{\rm Maj})^{-1}
(M_\nu^{\rm Dir})^T.}
In Eq.~\naivemat\ we estimated $M_\nu$ assuming that singlet neutrinos
play no role. Here we will prove that the same estimate applies also
to $M_\nu$ of Eq.~\Mlight\ if the conditions a-d hold.

If there are no holomorphic zeros in the neutrino mass matrices, then
we have
\eqn\naiMs{(M_{\nu_s}^{\rm Maj})_{ij}\sim M\lambda^{H(N_i)+H(N_j)},}
and
\eqn\naiMD{(M_{\nu}^{\rm Dir})_{ij}\sim v_u
\lambda^{H(L_i)+H(N_j)+H(\phi_u)}.}

We recall that for any non-singular $3\times3$ matrix,
\eqn\Msele{A=\pmatrix{a_{11}&a_{12}&a_{13}\cr a_{21}&a_{22}&a_{23}\cr
a_{31}&a_{32}&a_{33}\cr},}
we have
\eqn\Msmion{A^{-1}={1\over\det A}\pmatrix{a_{22}a_{33}-a_{23}a_{32}&
a_{13}a_{32}-a_{12}a_{33}&a_{12}a_{23}-a_{13}a_{22}\cr
a_{23}a_{31}-a_{21}a_{33}&
a_{11}a_{33}-a_{13}a_{31}&a_{13}a_{21}-a_{11}a_{23}\cr
a_{21}a_{32}-a_{22}a_{31}&
a_{12}a_{31}-a_{11}a_{32}&a_{11}a_{22}-a_{12}a_{21}\cr}.}
Taking into account that $M_{\nu_s}^{\rm Maj}$ is symmetric,
Eqs.~\naiMs\ and \Msmion\ lead straightforwardly to
\ref\GrNi{Y.~Grossman and Y.~Nir, Nucl. Phys. B448 (1995) 30,
hep-ph/9502418.}\
\eqn\naiMsmo{[(M_{\nu_s}^{\rm Maj})^{-1}]_{ij}\sim
             [(M_{\nu_s}^{\rm Maj})_{ij}]^{-1}\sim
             {1\over M}\lambda^{-H(N_i)-H(N_j)}.}
{}From Eqs.~\Mlight, \naiMD\ and \naiMsmo, we find:
\eqn\Mlij{(M_\nu)_{ij}\sim{v_u^2\over M}
\lambda^{H(L_i)+H(L_j)+2H(\phi_u)}.}

We find then that indeed \naivemat\ holds for this case in spite of
the presence of singlet neutrinos. As a result of \Mlij, we have
\eqn\numix{\sin\theta^\nu_{23}\sim1\ \Longrightarrow\ H(L_2)=H(L_3).}
Consequently, $\sin\theta^\ell_{23}\sim1$ is unavoidable, leading to
$(\delta^{LL}_{23})_{\rm eff}\sim1$. We conclude that, to suppress
$(\delta^{LL}_{23})_{\rm eff}$, the naive models have to be modified
beyond the addition of singlet neutrinos.

It is important to note that our proof here applies for any number of
singlet neutrinos. This is straightforward to see, using the
generalization of \Msmion\ to an $n_s\times n_s$ matrix.

\subsec{Holomorphic zeros}

We now consider again models with a horizontal $U(1)_1\times U(1)_2$
symmetry, where each of the $U(1)$ factors is broken by a single small
parameter, $\epsilon_1(-1,0)\sim \lambda^m$ and
$\epsilon_2(0,-1)\sim\lambda^n$.  In such models, neutrino mass
matrices with hierarchical masses (and weak mixing) are easily
achieved if $H_1(L_2)\neq H_1(L_3)$ and $H_2(L_2)\neq H_2(L_3)$.  If
in addition $L_2$ and $L_3$ have equal effective horizontal charge,
\eqn\Heff{H_{\rm eff}(L_2)=H_{\rm eff}(L_3),\quad 
(H_{\rm eff}=mH_1+nH_2),}
large 23 mixing in the charged lepton matrix can also be arranged.  In
Ref.~\GNS, this mechanism was employed in the framework of (2,0)
models. Such models then satisfy the condition of large mixing and
large hierarchy in the 23 neutrino sector, but, as discussed above,
also predict $\tau\ra \mu \gamma$ rates near the present bound.

We now consider $(2,n_s\geq2)$ models, that is, models where the
charges of at least two singlet neutrinos do affect the AN parameters.
In this case, the light neutrino mass matrix has the form of
Eq.~\Mlight.  We would like to obtain both large mixing and large
hierarchy from the neutrino matrix alone.  The exact conditions for
this to hold are complicated, but it is easy to show that a sufficient
condition is
\eqn\avohom{\eqalign{
(M_\nu^{\rm Dir})_{23} \sim (M_\nu^{\rm Dir})_{33} \gg&\
{\rm all\ other\ entries\ of\ } M_\nu^{\rm Dir} \cr
(M_{\nu_s}^{\rm Maj})^{-1}_{33} \gg&\
{\rm all\ other\ entries\ of\ } (M_{\nu_s}^{\rm Maj})^{-1}.}}
To reduce mixing in the charged lepton matrix, we may use 
holomorphy to produce
\eqn\avohol{(M_{\ell^\pm})_{23}=0,\ \ \ (M_{\ell^\pm})_{33}\neq0.}

As an example of how this mechanism works, we now present an explicit
(2,2) model. We take $m=n=1$, that is
$\epsilon_1\sim\epsilon_2\sim\lambda$, and assign the following set of
charges for the lepton fields:
\eqn\avhoch{L_2(-1,0),\ \ L_3(-2,1),\ \ N_2(0,0),\ \ N_3(2,0),\ \
\bar\ell_2(1,5),\ \ \bar\ell_3(5,-1).}
The $2\times2$ mass matrices in the 23 subspace have the following forms:
\eqn\avhomm{M_\nu^{\rm Dir}\sim v_u
\pmatrix{0&\lambda\cr0&\lambda\cr},\ \ \
M_{\nu_s}^{\rm Maj}\sim M
\pmatrix{1&\lambda^2\cr\lambda^2&\lambda^4\cr},\ \ \
M_{\ell^\pm}\sim v_d\pmatrix{\lambda^5&0\cr0&\lambda^3\cr}.}  These
matrices satisfy the conditions of Eqs.~\avohom\ and \avohol.  For the
mass ratios and mixing, we get the following estimates:
\eqn\avhoph{m(\nu_2)/m(\nu_3)=0,\ \
m(\ell^\pm_2)/m(\ell^\pm_3)\sim\lambda^2, \ \ V^\ell_{23}\sim1.}  (The
vanishing neutrino mass will be lifted when the first generation is
incorporated.) Note that the charged lepton mass matrix is diagonal in
the 23 subspace, and so the neutrino mixing is generated completely by
the neutrino mass matrix. Equation \avhomm\ leads to
\eqn\avhoK{(\delta^{LL}_{23})_{\rm eff}\sim\lambda^2,}
and $B(\tau\ra\mu\gamma)$ well below the bound.

\subsec{Discrete symmetries}

In models with a discrete $Z_p\times U(1)$ symmetry, it is possible to
enhance a light neutrino mass eigenvalue if some of the entries in the
neutrino mass matrix are larger than their would-be value if the
symmetry were continuous. In Ref.~\GNS, this mechanism was employed in
the framework of (2,0) models to build viable models of neutrino
parameters, where the large mixing must come from the charged lepton
sector.

In $(2,n_s\geq2)$ models one can also use the mechanism of discrete
symmetries to induce $V^\ell_{23}\sim1$ from the neutrino mass matrix
and not from the charged lepton mass matrix. In particular, we can
arrange for the conditions of Eq.~\avohom\ to hold again for the
neutrino mass matrix, while arranging for the discrete symmetry to
produce
\eqn\avdil{(M_{\ell^\pm})_{23}\ll(M_{\ell^\pm})_{33}.}

As an example, we now present an explicit (2,2) model. We take $p=5$,
that is a $Z_5\times U(1)$ symmetry, and $m=n=1$.  We assign the
following set of charges for the lepton fields:
\eqn\avdich{L_2(3,1),\ \ L_3(4,0),\ \ N_2(0,0),\ \ N_3(2,0),\ \
\bar\ell_2(2,4),\ \ \bar\ell_3(1,3).}
The $2\times2$ mass matrices in the 23 subspace have the following forms:
\eqn\avdimm{
M_\nu^{\rm Dir}\sim v_u
\pmatrix{\lambda^4&\lambda\cr\lambda^4&\lambda\cr},\ \ \
M_{\nu_s}^{\rm Maj}\sim M
\pmatrix{1&\lambda^2\cr\lambda^2&\lambda^4\cr},\ \ \
M_{\ell^\pm}\sim v_d
\pmatrix{\lambda^5&\lambda^8\cr\lambda^5&\lambda^3\cr},}
again satisfying Eq.~\avohom.  For the mass ratios and mixing, we get
the following estimates:
\eqn\avdiph{m(\nu_2)/m(\nu_3)\sim\lambda^{10},\ \
m(\ell^\pm_2)/m(\ell^\pm_3)\sim\lambda^2,\ \ V^\ell_{23}\sim1,} where
the source of the large neutrino mixing is the neutrino mass matrix.
Note that \avdil\ is satisfied, and Eq.~\avdimm\ leads to
\eqn\avhoK{(\delta^{LL}_{23})_{\rm eff}\sim\lambda^2}
and $B(\tau\ra\mu\gamma)$ well below the bound.

\subsec{Two breaking parameters}

Finally, we consider models in which a horizontal $U(1)$ symmetry is
broken by two parameters of opposite charge and equal magnitude
($\lambda(-1)$ and $\bar\lambda(+1)$).  In Ref.~\NiSh, viable models
that employ this mechanism with the condition \twbr\ were constructed
with large 23 mixing required to diagonalize the charged lepton
sector.

In $(2,n_s\geq2)$ models we can use the mechanism of two breaking
parameters to satisfy Eq.~\avohom\ by imposing
\eqn\avotbn{|H(L_2)+H(N_3)|=|H(L_3)+H(N_3)|.}
To suppress mixing in the charged lepton sector, it is sufficient to
note that in the generic case $H(N_3)\neq H(\bar\ell_3)$, and so
\twbr\ does not hold and consequently also \avdil\ is satisfied.

As an example of how this mechanism works, we present an explicit
(2,2) model.  We assign the following set of charges for the lepton
fields:
\eqn\avtbch{L_2(-1),\ \ L_3(-3),\ \ N_2(0),\ \ N_3(2),\ \
\bar\ell_2(-4),\ \ \bar\ell_3(6).}
The $2\times2$ mass matrices in the 23 subspace have the following forms:
\eqn\avtbmm{
M_\nu^{\rm Dir}\sim v_u
\pmatrix{\lambda&\lambda\cr\lambda^3&\lambda\cr},\ \ \
M_{\nu_s}^{\rm Maj}\sim M
\pmatrix{1&\lambda^2\cr\lambda^2&\lambda^4\cr},\ \ \
M_{\ell^\pm}\sim v_d
\pmatrix{\lambda^5&\lambda^5\cr\lambda^7&\lambda^3\cr}.}
The mass ratios and mixing are
\eqn\avtbph{m(\nu_2)/m(\nu_3)\sim\lambda^{4},\ \
m(\ell^\pm_2)/m(\ell^\pm_3)\sim\lambda^2,\ \ V^\ell_{23}\sim1.}  
where again, the ${\cal O}(1)$ neutrino mixing is from the neutrino
mass matrix.  Equation \avtbmm\ leads to
\eqn\avtbK{(\delta^{LL}_{23})_{\rm eff}\sim\lambda^2,}
and $B(\tau\ra\mu\gamma)$ well below the bound.

\subsec{Three Generation Models}

In the previous sections, we have constructed a variety of (2,2)
examples.  It is possible to extend these to (3,3) models that fit
both the AN and SN parameters and where the radiative charged lepton
decays are suppressed.  For example, in the two breaking parameter
framework, if we assign charges
\eqn\Hcha{L_1(+1),\ L_2(-3),\ L_3(-1),\
N_1(0),\ N_2(0),\ N_3(+2),\
\bar\ell_1(-6),\ \bar\ell_2(5),\ \bar\ell_3(1),}
we find $\Delta m^2_{12}/\Delta m^2_{23}\sim\lambda^8$,
$V_{12}^\ell\sim1$ and $V_{23}^\ell\sim1$, as appropriate for the AN
and for the VO solution of the SN. We also have $V_{e3}\ll1$,
consistent with CHOOZ and AN, and contributions to $\tau \ra \mu
\gamma$ and $\mu \ra e \gamma$ well below bounds.
We did not prove, however, that all the mechanisms discussed in this
section can be extended to the three generation case, nor have we
shown that all three of the SN solutions can be accommodated in such
models.

Finally, we note that it is possible to achieve the AN and SN
parameters \ANpar\ and \SNpar\ together with a suppressed
$(\delta^{LL}_{23})_{\rm eff}$ in models without singlet neutrinos but
with the horizontal charges of all three active neutrinos playing a
role in achieving the AN parameters.

We now give an explicit example of such a (3,0) model.  The horizontal
symmetry is $U(1)$, with two small breaking parameters, similar to the
previous subsection.  We assign the following set of charges for the
lepton fields:
\eqn\avtzch{L_1(-4),\ \ L_2(6),\ \ L_3(2),\ \ \bar\ell_1(9),\ \
\bar\ell_2(-4),\ \ \bar\ell_3(-2).}
The $3\times3$ mass matrices have the following forms:
\eqn\avtzmm{
M_{\nu_a}^{\rm Maj}\sim {v_u^2\over M}
\pmatrix{\lambda^8&\lambda^2&\lambda^2\cr\lambda^2&\lambda^{12}
 &\lambda^8\cr
\lambda^2&\lambda^8&\lambda^4\cr},\ \ \
M_{\ell^\pm}\sim v_d\pmatrix{\lambda^5&\lambda^8&\lambda^6\cr
\lambda^{15}&\lambda^2&\lambda^4\cr \lambda^{11}&\lambda^2&1\cr}.}
For the mass ratios and mixing, we get the following estimates:
\eqn\avtzph{\Delta m^2_{12}/\Delta m^2_{23}\sim\lambda^{4},\ \
m(\ell^\pm_2)/m(\ell^\pm_3)\sim\lambda^2,\ \ V^\ell_{23}\sim1.}
However, the mixing in the charged lepton matrix is suppressed, with
\eqn\avtbK{(\delta^{LL}_{23})_{\rm eff}\sim\lambda^4,\
(\delta^{RR}_{23})_{\rm eff}\sim\lambda^2,}
and $B(\tau\ra\mu\gamma)$ well below the bound.

\newsec{Conclusions}

We have studied the predictions of supersymmetric Abelian flavor
models for lepton flavor violating decays in view of measurements of
neutrino mass and mixing parameters. We have found no
model-independent predictions, and so are unable to conclude that
charged lepton flavor violating processes provide unambiguous tests of
this framework.  However, we can make the following interesting
observations:

\item{1.} For models without singlet neutrinos, a generic prediction
is that $B(\tau\ra\mu\gamma)$ is close to its current bound for
slepton masses of order 100 GeV and may be within reach of future
experiments.
\item{2.} Conversely, if slepton masses are found to be not too heavy
and the upper bound on $B(\tau\ra\mu\gamma)$ becomes stronger, then
typically rather complicated models, for example, those constructed
above involving singlet neutrinos in an essential way, are required
and the Abelian symmetry framework loses predictive power.
\item{3.} Many models where the solar neutrino problem is solved by
large mixing (either vacuum oscillations or large angle matter
enhanced oscillations) are excluded or strongly disfavored by current
bounds on $B(\mu\ra e\gamma)$ and $\mu-e$ conversion.
\item{4.} In models where the solar neutrino problem is
solved by small angle matter enhanced oscillations, current bounds on
$B(\mu\ra e\gamma)$ and $\mu-e$ conversion often already require
slepton masses to be above ${\cal O}(500\ {\rm GeV})$. In such models,
large signals are predicted in future experiments sensitive to $\mu
\ra e \gamma$ and $\mu-e$ conversion.  Given the projected
improvements of three to four orders of magnitude, such experiments
are extremely interesting and promising.
\item{5.} Conversely, if no signal appears in future experiments
probing $\mu \ra e \gamma$ and $\mu-e$ conversion, many models will be
excluded, and the framework of Abelian horizontal symmetries is again
typically required to become rather baroque and of limited predictive
power.
\item{6.} If experiments favor the vacuum oscillations solution with
near-maximal mixing, then Abelian flavor symmetries that lead to a
pseudo-Dirac structure in the 12 subspace of the neutrino mass matrix
provide an attractive explanation.

\vskip 3cm

\centerline{\bf Acknowledgments}

J.L.F. is supported by the Department of Energy under contract
No.~DE--FG02--90ER40542 and through the generosity of Frank and Peggy
Taplin. Y.N. is supported by the Department of Energy under contract
No.~DE--FG02--90ER40542, by the Ambrose Monell Foundation, by the
United States $-$ Israel Binational Science Foundation (BSF), by the
Israel Science Foundation founded by the Israel Academy of Sciences
and Humanities, and by the Minerva Foundation (Munich).  Y.S. is
supported in part by the Koret Foundation.

\listrefs
\end